# A Control Co-Design Framework to Achieve Solution Feasibility in Energy System Optimization Problems*

Tania Rifat Jahan and Donald J. Docimo, *Member, IEEE*

*Abstract*— This work explores methods to identify energy system designs for infeasible control co-design optimization problems. Control co-design, or CCD, has been recognized as a powerful tool to maximize energy system capabilities through simultaneous determination of plant and controller parameters. However, due to the inherent nonlinearities, complexity, and conflicting criteria of energy systems, CCD optimization problems are susceptible to infeasibility and can lack potential solutions. While transforming the optimization problem by relaxing constraints has been developed for optimal control infeasibility challenges, solution feasibility for CCD is relatively unexplored. This paper proposes a framework to convert infeasible optimization problems into solvable forms for a class of CCD problems. The framework introduces a procedure to rank metric bounds from least likely to most likely to cause infeasibility. This provides guidance to algorithmically relax a limited number of constraints, leaving others intact. The proposed framework is applied to a CCD problem for designing a battery within a microgrid. Comparison against a baseline approach for relaxing optimization problems shows the framework requires only a reduced number of iterations to determine a solution.

## I. INTRODUCTION

This paper explores the relationship between control co-design (CCD) optimization problems and solution feasibility. CCD, the coupled selection of plant and controller parameters, is an established area that enhances system performance beyond the capabilities of sequential plant design followed by optimal control [1], [2]. Such techniques have found value in energy systems. Combined plant and controller design has been implemented for thermal management systems [3], [4], unmanned vehicles [5], fuel cells [6], and wind energy generation [7], [8], [9], [10]. While the benefits vary based on the specifics of the problem, studies have shown over 50% improvements in energy system performance criteria by using CCD [11].

CCD is often expressed through an optimization problem described by the decision variables, objectives, and constraints. This structure enhances the ability to algorithmically identify local and global minima, but is not guaranteed to have a solution. A *feasible problem* is when one or more solutions exist that satisfy all equality and inequality constraints. On the other hand, an *infeasible problem* is unable to meet all constraints and does not have a solution. Infeasibility and failed design are recognized issues, with several methodologies developed to combat this. Traditionally, the constraints can be relaxed to convert a problem from infeasible to feasible, which can be related to some conversion of the problem into a multi-objective format [12], [13]. Exact penalty approaches prioritize constraint satisfaction over optimality [14], and primal-dual multi-objective optimization incorporates infeasible solutions alongside feasible solutions [15].

With respect to the intersection of design, control, and infeasible problems, studies often emphasize the controls side. As an example, one optimal control framework approximates the infeasible problem by minimizing the gap between constraint sets [16]. For energy system control, a push-and-pull co-evolutionary framework has been developed to identify infeasible solutions in scheduling of microgrid behavior [17]. However, there is limited work that explores infeasible CCD problems. A notable related study is centered on detection of infeasibility for co-design applied to hypersonic vehicles [18].

This paper aims to address the gap in the ability to generate solutions of infeasible CCD problems. The study targets a class of CCD problems with conflicting metrics are bound tightly enough to make the problem infeasible. This is handled through a proposed, iterative CCD framework that reformulates the optimization problem and algorithmically determines the subset of bounds to relax, transforming the problem from infeasible to feasible. Key to this framework is a ranking methodology that determines which metric bounds have a higher probability of leading to infeasibility. Emphasizing energy system CCD, this framework is applied to the control and design of a microgrid with domain-appropriate featured metrics.

The remainder of this paper is organized as follows. Section II defines the generalized CCD optimization problem of interest. Section III presents a specified problem for microgrid battery and controller design, providing an example of the challenges with managing infeasibility. Section IV develops the ranking-based CCD framework to transform and relax the optimization problem for obtaining a solution. Section V explores a more complex form of the energy system optimization problem, comparing the proposed framework against a baseline approach. This is used to develop insights into the impact of constraint relaxation on solutions of the optimization problem. Section VI ends the paper with a summary of the major elements and future steps.

*This material is based upon work supported by the National Science Foundation under Award No. 2324707.

Tania Rifat Jahan is with the Department of Mechanical and Aerospace Engineering, Texas Tech University, Lubbock, TX 79409 USA (e-mail: tjahan@ttu.edu).

Donald J. Docimo is with the Department of Mechanical and Aerospace Engineering, Texas Tech University, Lubbock, TX 79409 USA (phone: 806.834.4037; e-mail: donald.docimo@ttu.edu).

## II. PROBLEM STATEMENT

This section introduces the problem of interest regarding CCD problem infeasibility. Equation (1) presents the CCD problem. While setting objective function $J$ to zero is not required, this is imposed for simplification and formats the problem as a nonlinear constraint satisfaction problem. The decision variables to calculate, $\vec{\xi}$, are defined in (2). This includes $\vec{x}_k$ and $\vec{u}_k$, the state and input vectors, respectively, with time index $k = 1, \ldots, N$. This also includes $\vec{\theta}$ and $\vec{\varphi}$, the plant and controller design variables. The final decision variables are the $m = 1, \ldots, M$ featured metrics of interest, $\vec{\mu}$. A featured metric can represent critical criteria relating to the performance or limitations of the system. The constraints of (1), from top to bottom, include setting the initial states, $\vec{x}_{IC}$, the dynamics with nonlinear functions within $\vec{f}$, the control law with functions in $\vec{f}_c$, and the functions $g_m$ to calculate each $\mu_m$. The inputs, plant design variables, controller design variables, and metrics are all bound between lower and upper limits, as denoted by the subscripts $lb$ and $ub$, respectively.

$$\min_{\vec{\xi}} \quad J = 0$$

$$\begin{aligned}
\text{s.t.:} \quad & \vec{x}_1 = \vec{x}_{IC} \\
& \vec{x}_k = \vec{f}(\vec{x}_{k-1}, \vec{u}_{k-1}, \vec{\theta}), \quad \forall k \in [2, N] \\
& \vec{u}_k = \vec{f}_c(\vec{x}_k, \vec{x}_{k-1}, \vec{u}_{k-1}, \vec{\varphi}), \quad \forall k \in [1, N-1] \quad (1)\\
& \mu_m = g_m(\vec{x}_1, \ldots, \vec{x}_N, \vec{u}_1, \ldots, \vec{u}_{N-1}, \vec{\theta}, \vec{\varphi}), \forall m \in [1, M] \\
& \vec{u}_{lb,k} \leq \vec{u}_k \leq \vec{u}_{ub,k}, \quad \forall k \in [1, N-1] \\
& \vec{\theta}_{lb} \leq \vec{\theta} \leq \vec{\theta}_{ub} \\
& \vec{\varphi}_{lb} \leq \vec{\varphi} \leq \vec{\varphi}_{ub} \\
& \mu_{lb,m} \leq \mu_m \leq \mu_{ub,m}, \quad \forall m \in [1, M]
\end{aligned}$$

$$\vec{\xi} = [\vec{x}_1, \ldots, \vec{x}_N, \vec{u}_1, \ldots, \vec{u}_{N-1}, \vec{\theta}, \vec{\varphi}, \vec{\mu}] \quad (2)$$

Depending on the specific conditions and parameters, (1) can be an infeasible problem. In the case of infeasibility, the standard approach involves relaxation of a subset or all of the constraints [12]. This effectively converts constraint violations into objective terms that penalize larger violations. In the context of CCD, relaxation of (1)'s equality constraints, input bounds at each instance $k$, and plant and controller design variable bounds would alter the underlying system dynamics and ignore physical limitations. Therefore, relaxation is limited to the constraints bounding the featured metrics. To summarize the problem of interest:

*Equation (1) is a potentially infeasible optimization problem, due in part to bounds imposed on featured metrics $\vec{\mu}$. It is desired to relax the fewest number of metric bounds to alter the problem into that with a solution.*

This reflects conditions where it is more desirable to limit the number of constraint violations, rather than have all metric bounds violated by a small amount each.

## III. MOTIVATING EXAMPLE: MICROGRID CCD

This section explores a basic example of the CCD problem described in (1). To begin, an equivalent circuit model is defined to capture microgrid dynamics, and incorporated into the microgrid CCD problem. With the problem infeasible,

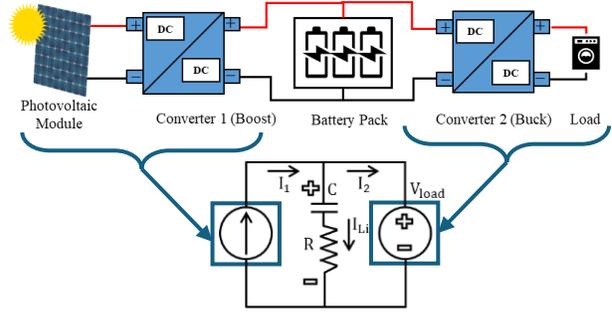

Figure 1. Microgrid plant configuration with the equivalent circuit representation.

different options for relaxing the optimization problem are explored.

### A. The Microgrid CCD Problem

To elucidate the problem of interest, a CCD optimization problem for microgrid battery sizing and control is considered. Fig. 1 presents the layout of the microgrid model alongside the equivalent circuit representation. For clarity, the circuit is kept simplistic without loss of generalization of the proposed methods. The system contains a battery pack, represented as a capacitor $C$ and resistor $R$ in series, connected in parallel to the remaining components. The photovoltaic (PV) panel and boost converter act as the energy generator, captured as current source $I_1$. The load is connected to the battery through a controllable buck converter, merged as controllable voltage source $V_{load}$.

The goal is to size the battery for given operating conditions. The battery dynamics are described by an ordinary differential equation (ODE), with state $x$ being the charge of the capacitor. Solving for the battery current $I_{Li}$, the continuous time state equation becomes:

$$\dot{x}(t) = -\frac{1}{RC}x(t) + \frac{1}{R}V_{load}(t) \quad (3)$$

Discretizing the model using the forward Euler method with time step $\Delta t$ provides the definition for $f$ of (1), with $\theta = 1 - \Delta t/(RC)$ and $u_k = (\Delta t/R)V_{load,k}$:

$$x_k = f(\cdot) = \theta x_{k-1} + u_{k-1} \quad (4)$$

Note that vector arrows have been dropped due to the state and input being scalar variables. The initial condition is set to $x_{IC} = 10$ for this example. The controller function $f_c$ uses proportional state feedback with reference state $x_r = 20$ and gain as the decision variable $\varphi$:

$$u_k = f_c(\cdot) = \varphi(x_r - x_k) \quad (5)$$

To start, two featured metrics are defined. The first represents an equivalent to total state tracking error with a terminal cost, and the second represents total control effort:

$$\mu_1 = g_1(\cdot) = x_1 + \cdots + x_{N-1} - x_N \quad (6)$$

$$\mu_2 = g_2(\cdot) = 5(u_1 + \cdots + u_{N-1}) \quad (7)$$

Here, bounds for the decision variables include $u_{lb} = 0$, $u_{ub} = 7$, $\theta_{lb} = 0.1$, $\theta_{ub} = 0.5$, $\varphi_{lb} = 0.1$, $\varphi_{ub} = 0.5$, $\mu_{lb,1} = 0$, $\mu_{ub,1} = 8$, $\mu_{lb,2} = 0$, and $\mu_{ub,2} = 20$. With these bounds, the metrics conflict: as $\theta$ increases, $\mu_1$ increases and $\mu_2$ decreases. If $\varphi$ increases, the reverse occurs. With $N = 3$, these conditions together define (1), which is solved using the MATLAB function `fmincon`.

## B. Problem Relaxation

For the conditions described, the optimization problem does not have a feasible solution. Any choice of the design variables $\theta$ and $\varphi$, which in turn will determine the input, state, and metric values through the equality constraints, will cause $\mu_1$ or $\mu_2$ to exceed a limit. To overcome this, a sacrifice must be made, and a subset of the constraints of the optimization problem must be relaxed. This will be accomplished by defining slack variables $s_1$ and $s_2$ that correspond to the bounds of $\mu_1$ and $\mu_2$, respectively. Slack variables indicate how much a metric exceeds its bound, using (8) for $m = [1,2]$ to replace the metric bound inequality constraints within (1).

$$\begin{aligned} s_m &\geq 0 \\ -\mu_m - s_m &\leq -\mu_{lb,m} \\ \mu_m - s_m &\leq \mu_{ub,m} \end{aligned} \quad (8)$$

The new constraints can be used to replace bounds of the constraints mentioned earlier. If both sets of metric bounds are relaxed, the cost function would be redefined as:

$$J = w_1 s_1^2 + w_2 s_2^2 \quad (9)$$

where $w_1$ and $w_2$ are the weights associated with each metric. It should be noted that if only one metric's constraint is relaxed, only that equivalent term would appear in the cost function (9).

There are three options for relaxing the problem. First, the baseline approach is to relax all metric bounds. This will depend heavily on the weights selected: as shown in Fig. 2(a), a choice of $w_1 = 0.4$ and $w_2 = 0.6$ will yield $\mu_2$ to be nearly within its bounds. However, as shown by Fig. 2(b), a choice of $w_1 = 0.999$ and $w_2 = 0.001$ will set both $\mu_1$ and $\mu_2$ to be well outside their upper limits. This does not meet the desired outcome as expressed by the problem of interest in Section II. The second approach is to relax $\mu_1$ and keep $\mu_2$ constrained. As shown in Fig. 2(c), this yields a desired solution, with one metric kept within its bounds. The third approach relaxes $\mu_2$ and keeps $\mu_1$ constrained, but this still produces an infeasible problem. The latter two approaches are subsets of the first, however, haphazard weight selection can lead to both bounds being violated or excessively high violations of one metric. Guidance is required to determine which metric(s) to select for relaxing the bounds, as this knowledge is rarely known a priori.

## IV. A CCD FRAMEWORK FOR PROBLEM SOLUTION FEASIBILITY

This section presents a framework to restructure a CCD problem to have solution feasibility by relaxing a limited number of featured metric bounds. The previous section showed that it is possible to relax only certain metrics for some infeasible problems, but did not provide knowledge of how to select which constraints to relax. The proposed framework provides this using a ranking-based testing step to determine the metrics most likely to violate their bounds. Fig. 3 presents a flowchart of the steps of the framework.

### 2.1. Step 1: Define the Original CCD Problem

This first step serves to set up the CCD problem as an optimization problem. In the context of this work, the problem is limited to the constraint satisfaction problem described by (1). The problem is potentially infeasible, and the only acceptable constraints to be relaxed are those describing the bounds of the featured metrics.

### 2.2. Step 2: Reformulate the Optimization Problem

The second step is to reformulate the structure of the optimization problem to permit relaxing the metric bounds. Equation (1) is reformatted into (10), with four major changes: (i) Slack variables $\vec{s}$ are added to the decision variables $\vec{\xi}$, with one slack variable per featured metric. (ii) The objective function is modified to be the sum of each weight $w_m > 0$ multiplied by the corresponding slack variable squared. Targeted selection of each weight is not critical to meet the desired requirements described by the problem of interest, which is concerned primarily with how many metrics exceed their bounds. (iii) The constraints for the metric bounds are rewritten to incorporate the slack variables. (iv) The selection variables $\vec{z}$ are introduced to the problem. Each $z_m$ is binary: a value of 0 relaxes the corresponding metric, and a value of 1 keeps the metric bounded by hard constraints.

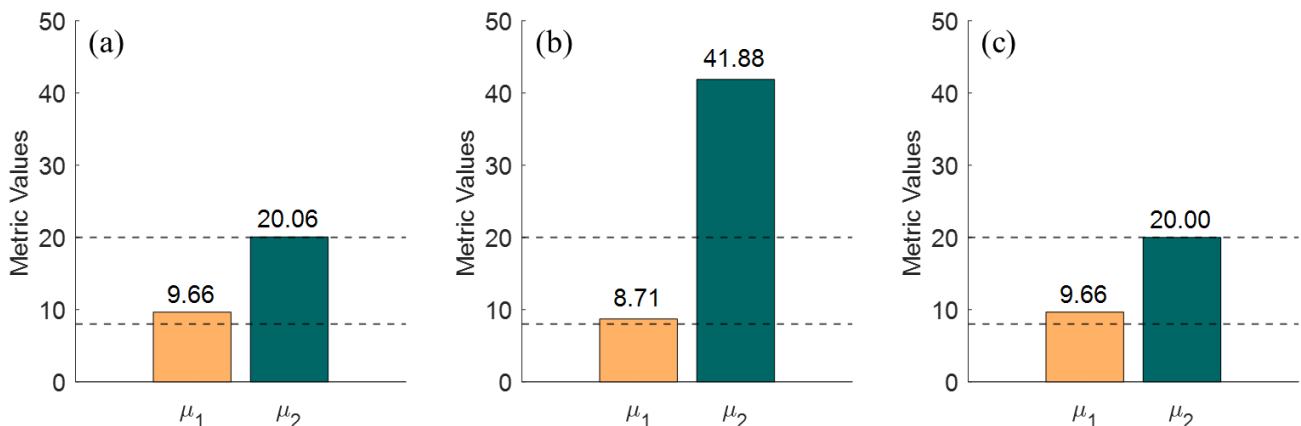

Figure 2. Metric values for the optimal solutions of the baseline approach with (a) desirable and (b) undesirable weights, and (c) relaxing only the first metric. The dashed lines represent the upper bounds.

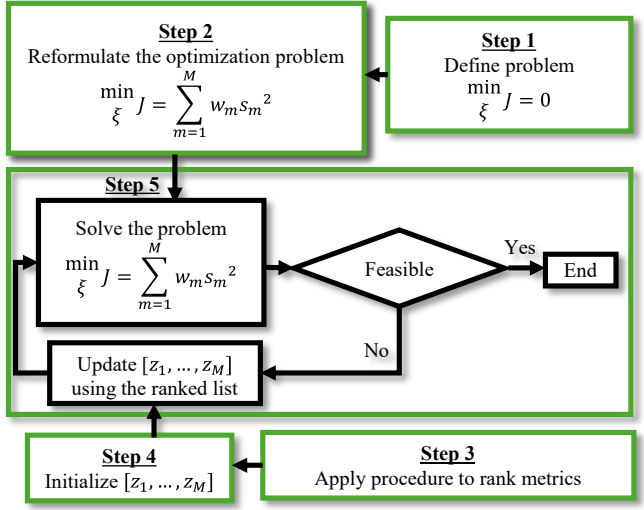

Figure 3. Flowchart for the ranking-based CCD framework to support problem feasibility.

$$\min_{\vec{\xi}} \quad J = \sum_{m=1}^{M} w_m s_m^2$$

s.t.: $\vec{x}_1 = \vec{x}_{IC}$
$\vec{x}_k = \vec{f}(\vec{x}_{k-1}, \vec{u}_{k-1}, \vec{\theta}), \quad \forall\, k \in [2, N]$
$\vec{u}_k = \vec{f}_c(\vec{x}_k, \vec{x}_{k-1}, \vec{u}_{k-1}, \vec{\varphi}), \quad \forall\, k \in [1, N-1]$
$\mu_m = g_m(\vec{x}_1, \ldots, \vec{x}_N, \vec{u}_1, \ldots, \vec{u}_{N-1}, \vec{\theta}, \vec{\varphi}), \forall\, m \in [1, M]$ (10)
$\vec{u}_{lb,k} \le \vec{u}_k \le \vec{u}_{ub,k}, \quad \forall\, k \in [1, N-1]$
$\vec{\theta}_{lb} \le \vec{\theta} \le \vec{\theta}_{ub}$
$\vec{\varphi}_{lb} \le \vec{\varphi} \le \vec{\varphi}_{ub}$
$s_m \ge 0, \quad \forall\, m \in [1, M]$
$-\mu_m - (1 - z_m) s_m \le -\mu_{m_{lb}}, \quad \forall\, m \in [1, M]$
$\mu_m - (1 - z_m) s_m \le \mu_{m_{ub}}, \quad \forall\, m \in [1, M]$

## 2.3. Step 3: Rank the Featured Metrics

To inform the selection of values for each $z_m$ in later steps, a ranking procedure is developed to determine which metrics should be relaxed earlier in the solution process. This work prioritizes first relaxing metrics that are more likely to exceed their bounds. The ranking is determined using a method akin to a short grid search:

1. Firstly, a small population of candidate designs is defined by selecting values for the plant and controller design variables between the corresponding bounds for $\vec{\theta}$ and $\vec{\varphi}$. Alternatively, values can be selected randomly from a uniform distribution.
2. Then, the states, inputs, and featured metrics are calculated for each candidate design.
3. For each design, each metric is compared against its bounds.
4. Across all designs, the total number of violations each metric has is summed.
5. Finally, the metrics are ranked. The one with the most violations across the designs is ranked last, as it is prone to violate its bounds. The one with the least violations is ranked first.

The ranking of metrics, from least to most violations, is influenced by the outcome of the motivating example. This provides a list of which metric bounds to relax, with the lowest ranked one the most likely to promote infeasibility.

## 2.4. Step 4: Initialize the Selection Variables

Before attempting to find a solution with the restructured optimization problem, initialization of $\vec{z} = [z_1, \ldots, z_M]$ is required. One choice is to set all to 1 to initially look at the problem with hard constraints. However, the previous step's information can be used to inform the selected values corresponding to each metric.

## 2.5. Step 5: Algorithmically Solve the Reformatted Problem

The last step of the framework is to solve the reformatted CCD problem with an outer loop for updating $z$ values, as shown in Fig. 3. The flow of the algorithm is as follows:

1. Input the initialized values for $\vec{z}$ from step 4.
2. Apply a nonlinear optimization problem solver to (10).
3. If the problem is found to still be infeasible, update $\vec{z}$ by moving up the ranked list by at least one row, and setting all values below to 0. Return to 2.
4. If the problem is feasible, at least one solution has been identified and the algorithm ends.

## V. MICROGRID EXAMPLE ON PROBLEM SOLUTION FEASIBILITY

This section applies the proposed framework to the microgrid CCD problem and compares the results against a baseline approach. The example from Section II.A is extended to include two additional metrics:

$$\mu_3 = \theta \quad (11)$$

$$\mu_4 = \frac{1}{\theta} \quad (12)$$

The third metric $\mu_3$ is associated with degradation of the battery pack: increasing $\theta$ is the equivalent to decreasing the number of cells in the pack. The workload of each cell increases, increasing the degradation of the pack. The fourth metric $\mu_4$ is representative of battery pack mass. Alternatively, this is also representative of greenhouse gas (GHG) emissions created during the manufacturing of the battery pack [19]. The literature reveals the relationship between manufacturing of batteries and emission of $CO_2$ equivalents into the environment [20]. The bounds of the two new metrics are $\mu_{lb,3} = 0$, $\mu_{ub,3} = 0.1$, $\mu_{lb,4} = 0$, and $\mu_{ub,4} = 5$.

### A. Application of the CCD Framework

Applying the first two steps of the framework is straightforward. For step 1, the optimization problem of (1) for this microgrid has been previously defined. For step 2, the conversion to (10) requires the definition of the $M = 4$ slack variables and selection variables. The weights are selected to all be one. Step 3 requires testing of a small population of values – varying $\theta$ and holding $\varphi$ constant at 0.1, the results of these tests are presented in Table I. From these tests, $\mu_1$ violates its constraints the most, followed by $\mu_3$, and then $\mu_2$ and $\mu_4$. This reveals the ranking, from the lowest ($\mu_1$) being first to be relaxed, to the highest ($\mu_2$ and $\mu_4$) that will only be relaxed if absolutely necessary. Step 4 is to initialize $\vec{z}$, which is selected to be all ones.

TABLE I. RANKING OF THE METRICS BASED ON CONSTRAINT VIOLATIONS (VIOLATIONS HIGHLIGHTED IN RED).

| Test | $\theta$ | $\mu_1$ | $\mu_2$ | $\mu_3$ | $\mu_4$ |
|---|---|---|---|---|---|
| 1 | 0.1 | 10.0 | 14.0 | 0.1 | 10.0 |
| 2 | 0.2 | 10.7 | 13.5 | 0.2 | 5.00 |
| 3 | 0.3 | 11.2 | 13.0 | 0.3 | 3.33 |
| 4 | 0.4 | 11.5 | 12.5 | 0.4 | 2.50 |
| 5 | 0.5 | 11.6 | 12.0 | 0.5 | 2.00 |

Enacting on the procedure for step 5 yields the following: with the selected initial values of $\vec{z}$, the problem is infeasible. The ranking of the metrics is consulted – moving up one row in the ranking indicates setting $z_1 = 0$ for the next attempt. Once again, a solution to the optimization problem does not exist, and the ranking is consulted again. This time, $z_1 = 0$ and $z_3 = 0$, indicating both $\mu_1$ and $\mu_3$ bounds should be relaxed. With this selection for $\vec{z}$, the problem becomes feasible and a solution is identified. Fig. 4 presents the value of each featured metric for the identified solution. Two metrics, $\mu_2$ and $\mu_4$, are within their bounds. This indicates that the microgrid CCD problem requires two metrics to be relaxed to determine a solution.

## B. Comparison Against a Baseline Approach

To determine if the proposed framework relaxes a minimal number of metric bounds, comparison against a baseline approach is performed. The baseline approach is to check whether (1) is infeasible – if it is, all metric bounds are relaxed (i.e., all $\vec{z}$ values in (10) are set to zero). With all bounds relaxed, the selection of proper weights becomes imperative. To reflect the potential options, each weight can be 0, 0.25, 0.75, or 1. All combinations are swept through, yielding 256 trials in total.

Fig. 5 presents the number of metrics within their bounds per trial of the baseline approach. Table II provides a detailed numerical snapshot of this data for 10 representative trials, highlighting the metrics within their bounds. Fig. 6 presents a plot of the metrics for trial 50, showing that some designs keep two metrics well under the defined upper bounds. From this data, two insights are obtained:

*Insight #1:* The proposed framework and best trials from the baseline approach identify the same number of metrics to relax. The framework identifies that two metric bounds must be relaxed to produce feasible solutions, $\mu_1$ and $\mu_3$. Fig. 5 shows that the baseline never has more than two metrics within its bounds. This is indicative that a minimum of two metrics must be relaxed to identify a solution to the specified CCD problem. It is a promising result that shows the proposed framework is able to relax the proper number of metric bounds.

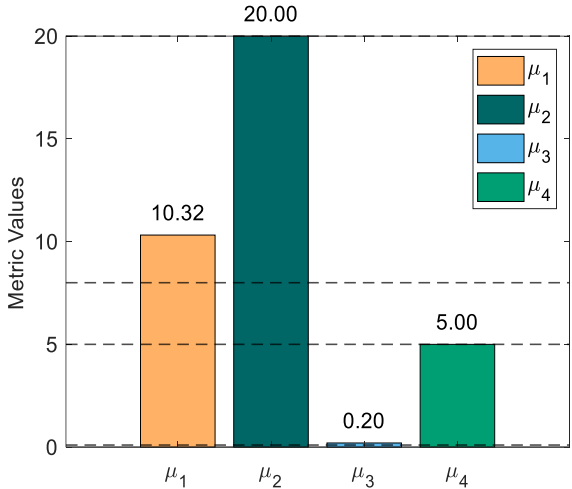

Figure 4. Metric values as determined by the proposed framework.

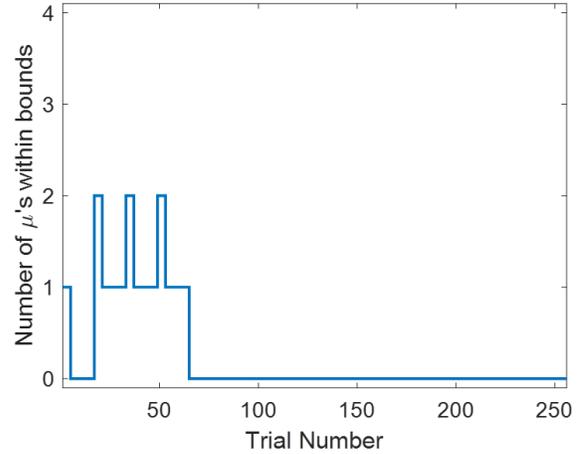

Figure 5. Number of metrics that satisfy the original hard constraints per trial of the baseline approach.

Table II. SAMPLE DATA FROM THE BASELINE APPROACH TRIALS, WITH BLUE HIGHLIGHTS INDICATING THE METRIC IS WITHIN ITS BOUNDS.

| Trial | $w_1$ | $w_2$ | $w_3$ | $w_4$ | $\mu_1$ | $\mu_2$ | $\mu_3$ | $\mu_4$ | Number of $\mu$'s within their bounds |
|---|---|---|---|---|---|---|---|---|---|
| 1 | 0 | 0 | 0 | 0 | 10.20 | 31.85 | 0.3089 | **3.237** | 1 |
| 5 | 0 | 0 | 0.25 | 0 | 8.791 | 39.30 | 0.1001 | 9.986 | 0 |
| 16 | 0 | 0 | 1 | 1 | 9.388 | 40.76 | 0.1998 | 5.004 | 0 |
| 50 | 0 | 1 | 0 | 0.25 | 11.13 | **16.82** | 0.3554 | **2.814** | 2 |
| 64 | 0 | 1 | 1 | 1 | 10.47 | **17.34** | 0.1998 | 5.004 | 1 |
| 66 | 0.25 | 0 | 0 | 0.25 | 9.045 | 57.71 | 0.1910 | 5.236 | 0 |
| 88 | 0.25 | 0.25 | 0.25 | 1 | 10.28 | 20.13 | 0.1951 | 5.124 | 0 |
| 100 | 0.25 | 0.75 | 0 | 1 | 10.29 | 20.04 | 0.1952 | 5.123 | 0 |
| 134 | 0.75 | 0 | 0.25 | 0.25 | 8.967 | 58.03 | 0.1786 | 5.598 | 0 |
| 168 | 0.75 | 0.75 | 0.25 | 1 | 10.24 | 20.12 | 0.1871 | 5.344 | 0 |

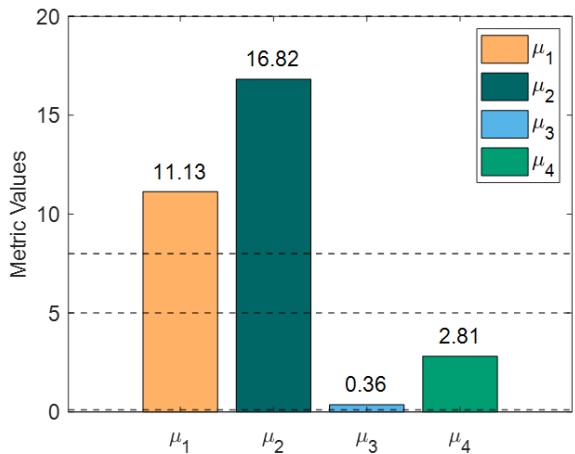

Figure 6. Metric values for the baseline approach, trial 50.

*Insight #2:* The proposed framework is able to generate a feasible version of the optimization problem faster than the median of the baseline approach. After determining the original problem to be infeasible, the proposed framework requires two attempts to solve the optimization problem: one with $\mu_1$ relaxed, and another with both $\mu_1$ and $\mu_3$ relaxed. On the other hand, Fig. 5 shows that only 12 of the 256 trials evaluated for the baseline approach identified a design with two metrics within bounds. This is under 4.7% of the trials. Assuming random picking of the weights to test, the baseline approach has a 90.8% chance of requiring more than the two attempts the proposed framework necessitates. As the problem of interest is to limit the number of constraint violations, rather than permitting all metric bounds to be violated by small amounts, comparing the magnitude of the metric values is beyond the scope of this study. While iterative tuning for the weights can be implemented for the baseline approach, most initial guesses of the weights will require more evaluations for the baseline approach as compared to the proposed framework. A baseline approach with weight tuning will scale unfavorably for optimization problems with more metrics.

## VI. CONCLUSION

This paper develops a CCD framework to attain solutions for infeasible, constrained problems. The framework reformulates the optimization problem using slack variables to relax a subset of bounds for featured metrics of the system. With the desired outcome being the relaxation of the fewest metric bounds permittable, a ranking method is incorporated to identify which constraints to prioritize relaxing and which constraints to retain. Tested on a battery microgrid CCD problem, the framework requires two iterations to determine a design for the plant and controller, while the probability of the baseline approach matching this is under 10%. Comparison with the baseline further indicates that the framework has the potential of solving an infeasible CCD problem with fewer constraint violations. Future work will explore multi-level optimization structures to incorporate deeper searches using the developed ranking method for large-scale problems. Another potential extension of this work will incorporate nonzero, mandatory objective functions alongside the constraint satisfaction goal.

REFERENCES

[1] H. K. Fathy, J. A. Reyer, P. Y. Papalambros, and A. G. Ulsoy, "On the coupling between the plant and controller optimization problems," in *Proceedings of the American Control Conference*, 2001, pp. 1864–1869.
[2] M. Garcia-Sanz, "Control Co-Design: An engineering game changer," *Advanced Control for Applications: Engineering and Industrial Systems*, vol. 1, no. 1, Dec. 2019.
[3] A. L. Nash and N. Jain, "Combined Plant and Control Co-Design for Robust Disturbance Rejection in Thermal-Fluid Systems," *IEEE Transactions on Control Systems Technology*, vol. 28, no. 6, pp. 2532–2539, 2019.
[4] C. T. Aksland and A. G. Alleyne, "An Approach to Simultaneous Topology, Parametric, and Feedback Control Co-design: Applications to Conservation-Based Energy Systems," *IEEE Control Syst.*, vol. 45, no. 3, pp. 28–55, 2025.
[5] C. T. Aksland, D. L. Clark, C. A. Lupp, and A. G. Alleyne, "Closed-Loop Control and Plant Co-Design of a Hybrid Electric Unmanned Air Vehicle," *Journal of Dynamic Systems, Measurement and Control, Transactions of the ASME*, vol. 146, no. 1, Jan. 2024.
[6] M. J. Kim and H. Peng, "Power management and design optimization of fuel cell/battery hybrid vehicles," *J. Power Sources*, vol. 165, no. 2, pp. 819–832, Mar. 2007.
[7] L. Y. Pao, M. Pusch, and D. S. Zalkind, "Control Co-Design of Wind Turbines," *Annu. Rev. Control Robot. Auton. Syst.*, vol. 7, p. 17, 2024.
[8] X. Du, L. Burlion, and O. Bilgen, "Control Co-Design for Rotor Blades of Floating Offshore Wind Turbines," in *ASME International Mechanical Engineering Congress and Exposition*, 2020.
[9] T. Cui, J. T. Allison, and P. Wang, "Reliability-based control co-design of horizontal axis wind turbines," *Structural and Multidisciplinary Optimization*, vol. 64, no. 6, pp. 3653–3679, Dec. 2021.
[10] H. Sharma, W. Wang, B. Huang, T. Ramachandran, and V. Adetola, "Multi-Objective Control Co-design Using Graph Based Optimization for Offshore Wind Farm Grid Integration," IEEE, 2024.
[11] A. Bhattacharya, S. Vasisht, V. Adetola, S. Huang, H. Sharma, and D. L. Vrabie, "Control co-design of commercial building chiller plant using Bayesian optimization," *Energy Build.*, vol. 246, Sep. 2021.
[12] Z. Cai and Y. Wang, "A multiobjective optimization-based evolutionary algorithm for constrained optimization," *IEEE Transactions on Evolutionary Computation*, vol. 10, no. 6, pp. 658–675, 2006.
[13] S. Venkatraman and G. G. Yen, "A generic framework for constrained optimization using genetic algorithms," *IEEE Transactions on Evolutionary Computation*, vol. 9, no. 4, pp. 424–435, Aug. 2005.
[14] R. Byrd, Frank. Curtis, and J. Nocedal, "Infeasibility Detection and SQP Methods for Nonlinear Optimization," *SIAM Journal on Optimization*, vol. 20, no. 5, pp. 2281–2299, 2010.
[15] T. Hanne, "On utilizing infeasibility in multiobjective evolutionary algorithms," Springer, 2009, pp. 113–122.
[16] R. S. Burachik, C. Y. Kaya, and W. M. Moursi, "Infeasible and Critically Feasible Optimal Control," *J. Optim. Theory Appl.*, vol. 203, no. 2, pp. 1219–1245, 2024.
[17] Y. Hua, B. Liao, H. Zhang, L. Liu, Y. Li, and Y. Yang, "Optimal Scheduling of Microgrid Using Constrained Multi-Objective Optimization with Additional Objectives," in *2024 6th International Conference on Data-driven Optimization of Complex Systems (DOCS)*, 2024, pp. 501–506.
[18] C. Van Der Heide, P. Cudmore, I. Jahn, V. Bone, P. M. Dower, and C. Manzie, "Feasibility Detection for Nested Codesign of Hypersonic Vehicles," in *2023 62nd IEEE Conference on Decision and Control (CDC)*, IEEE, 2023, pp. 632–637.
[19] T. R. Jahan, A. S. Ouedraogo, and D. J. Docimo, "A Study on Control Co-Design for Optimizing Microgrid Sustainability," *IFAC-PapersOnLine*, vol. 58, no. 28, pp. 636–641, 2024.
[20] H. Hao, Z. Mu, S. Jiang, Z. Liu, and F. Zhao, "GHG Emissions from the production of lithium-ion batteries for electric vehicles in China," *Sustainability*, vol. 9, no. 4, p. 504, 2017.